\documentclass[journal,onecolumn]{IEEEtran}
\usepackage{amssymb}
\usepackage{caption2}
\usepackage{amsmath}
\usepackage{multirow}
\usepackage{amsthm}
\usepackage[ruled,vlined,commentsnumbered]{algorithm2e}
\newtheorem{theorem}{Theorem}[section]
\newtheorem{definition}[theorem]{Definition}
\newtheorem{lemma}[theorem]{Lemma}
\newtheorem{corollary}[theorem]{Corollary}

\newtheorem{exmp}[theorem]{Example}

\begin{document}

\title{Strongly Separable Codes}
\author{Jing Jiang, ~Minquan Cheng and Ying Miao
\thanks{The research of Cheng was supported in part by NSFC (No.~11301098),
Guangxi  Natural Science Foundations (No.~2013GXNSFCA019001), Foundation
of Guangxi Education Department (No.~2013YB039), and Program on the High Level
Innovation Team and Outstanding Scholars in Universities of Guangxi Province.
The research of Y. Miao was supported by JSPS Grant-in-Aid for Scientific Research (C) under Grant No.~24540111.
}

\thanks{J. Jiang and Y. Miao are with the Department of Social Systems and Management,
Graduate School of Systems and Information Engineering, University of Tsukuba, Tsukuba 305-8573, Japan.
E-mails: jjiang2008@hotmail.com, miao@sk.tsukuba.ac.jp.}

\thanks{M. Cheng is with the School of Mathematics and Statistics, Guangxi Normal University, Guilin 541004, P. R. China.
E-mails: chengqinshi@hotmail.com,  dhwu@gxnu.edu.cn.}

}
\maketitle

\begin{abstract}
Binary $t$-frameproof codes ($t$-FPCs) are used in multimedia fingerprinting schemes
where the identification of authorized users taking part in the averaging collusion
attack is required.  In this paper, a binary strongly $\overline{t}$-separable
code ($\overline{t}$-SSC) is introduced to improve such a scheme based on a binary $t$-FPC.
A binary $\overline{t}$-SSC has the same traceability as a binary $t$-FPC but has more codewords
than a binary $t$-FPC. A composition construction for binary $\overline{t}$-SSCs from $q$-ary $\overline{t}$-SSCs
is described, which stimulates the research on $q$-ary $\overline{t}$-SSCs with short length. Several infinite series of
optimal $q$-ary $\overline{2}$-SSCs of length $2$ are derived from the fact that a $q$-ary $\overline{2}$-SSC of
length $2$ is equivalent to a $q$-ary $\overline{2}$-separable code of length $2$.
Combinatorial properties of $q$-ary $\overline{2}$-SSCs of length $3$ are investigated,
and a construction for $q$-ary $\overline{2}$-SSCs of length $3$ is provided. These
$\overline{2}$-SSCs of length $3$ have more than $12.5\%$ codewords than $2$-FPCs of length $3$ could have.
\end{abstract}


\section{Introduction}
\label{intro}

Multimedia content protection has become an important security issue
in recent years, as illegal redistribution of licensed materials
has become increasingly prevalent. Cryptographic techniques were introduced to ensure that
only authorized users are able to use them. Unfortunately, cryptographic approaches
are limited in that once the content is decrypted,
it can potentially be copied and redistributed freely.
In order to hinder the unauthorized redistribution of digital data, digital fingerprinting was introduced
to trace the authorized users who redistribute their contents for unintended purposes \cite{BS}.
Fingerprints for multimedia data can be embedded through a variety of watermarking techniques
prior to their authorized distribution \cite{CKLS,PZ}.

Fingerprinting can be an effective technique for inhibiting malicious
authorized users from distributing their copies of the
media. However, if fingerprints are not well designed, fingerprinting systems would become invalid when
a group of users form a collusion by cleverly combining their copies to
create a pirate copy. Among others, the averaging attack  is an attempt to remove the
embedded fingerprints by averaging all the fingerprinted signals
with an equal weight for each colluder, so that no colluder would
take more of a risk than any other colluders. This attack also reduces the
power of each contributing fingerprint and makes the colluded
signal have better perceptual quality.

Various anti-collusion codes have been introduced to resist collusion attacks.
Frameproof codes (FPCs) were first introduced to prevent a coalition from framing a user not in the coalition
in \cite{BS}, but widely considered as having no traceability for generic
digital data (see for example \cite{STW}).
However, Cheng and Miao \cite{CM} showed that frameproof codes actually have traceability for multimedia contents.
This greatly strengthens the importance of frameproof codes in fingerprinting.
Unfortunately, in most cases, the number of codewords in a $t$-FPC is too small to be of practical use.
In this paper,
we  introduce a new notion of a binary strongly $\overline{t}$-separable code ($\overline{t}$-SSC) which has weaker
requirements than a binary $t$-FPC but has the same traceability as a binary $t$-FPC.
Usually, $\overline{t}$-SSCs have much more codewords than $t$-FPCs could have.

This paper is organized as follows. In Section \ref{pre}, we first recall some basic concepts of
fingerprinting, collusion, and detection. In Section \ref{FPCSSC}, we introduce the concept of an
SSC, and describe a colluder-tracing algorithm based on a binary SSC. We also show a composition construction
for binary SSCs from $q$-ary SSCs, which makes the study of $q$-ary SSCs with short length important.
In Section \ref{constr}, we derive several infinite series of optimal $q$-ary $\overline{2}$-SSCs of length $2$
from the fact that a $q$-ary $\overline{2}$-SSC of
length $2$ is equivalent to a $q$-ary $\overline{2}$-separable code ($\overline{2}$-SC) of length $2$.
Combinatorial properties of $q$-ary $\overline{2}$-SSCs of length $3$ are also investigated and a construction
for $q$-ary $\overline{2}$-SSCs of length $3$ is also presented in Section \ref{constr}.
Finally, conclusions will be given in Section \ref{conclu}.

%%%%%%%%%%%%%%%%%%%%%%%%%%%%
\section{Preliminaries } %
\label{pre}                                                                  %
%%%%%%%%%%%%%%%%%%%%%%%%%%%

In this section, we give a brief review on some basic terminologies.
The interested reader is referred to \cite{CM,LTWWZ} for more detailed information.

In collusion-resistant fingerprinting, we want to design fingerprints which can be used to
trace and identify colluders after collusion attacks, together with robust embedding of
fingerprints into multimedia host signals. Spread-spectrum additive embedding is one of
the widely employed robust embedding techniques.
Let ${\bf x}$ be the host multimedia signal, $\{ {\bf u}_i \ | \ 1 \leq i \leq n\}$ be an
orthonormal basis of noise-like signals,  $\{{\bf w}_j = ({\bf w}_j(1), {\bf w}_j(2),
\ldots, {\bf w}_j(n)) = \sum_{i=1}^{n}b_{ij}{\bf u}_i\ | \ 1 \leq j \leq M\}$, $b_{ij} \in \{0,1\}$,
be a family of watermarks, and $\{ \alpha{\bf w}_j \ | \ 1 \leq j \leq M\}$, $\alpha \in \mathbb{R}^{+}$, be the
scaled watermarks to achieve the imperceptibility as well as to control the energy of the embedded watermark.
The watermarked version of the content ${\bf y}_j = {\bf x} + \alpha{\bf w}_j$, $1 \leq j \leq M$, is then assigned to
the authorized user $U_j$ who has purchased the rights to access ${\bf x}$.
The fingerprint ${\bf w}_j$  assigned to $U_j$ can be represented uniquely by a vector (called codeword)
${\bf b}_j = (b_{1j}, b_{2j}, \ldots, b_{nj})^{T} \in \{0,1\}^{n}$ because of the linear independence
of the basis $\{ {\bf u}_i \ | \ 1 \leq i \leq n\}$.

When $t$ authorized users, say $U_{j_1}, U_{j_2}, \ldots, U_{j_t}$,
who have the same host content  but distinct fingerprints come together,
we assume that they have no way of manipulating the individual orthonormal signals,
that is, the underlying codeword needs to be taken and proceeded as a single entity,
but they can carry on a linear collusion attack to generate a pirate copy from their $t$ fingerprinted contents,
so that the venture traced by the pirate copy can be attenuated. In additive embedding,
this is done by linearly combining the $t$ fingerprinted contents
$\sum_{\ell=1}^{t}\lambda_{j_{\ell}}{\bf y}_{j_{\ell}}$, where the weights $\{{\lambda}_{j_{\ell}} \in  \mathbb{R}^{+} \ | \ 1 \leq \ell \leq t\}$
satisfy the condition $\sum_{{\ell}=1}^{t}\lambda_{j_{\ell}} = 1$ to maintain the average intensity of the original multimedia signal.
In this case, the energy of each of the scaled watermarks $\alpha{\bf w}_{j_{\ell}}$ is reduced by a factor of $\lambda_{j_{\ell}}^{2}$,
therefore, the trace of $U_{j_{\ell}}$'s fingerprint becomes weaker and thus $U_{j_{\ell}}$ is less likely to be caught by the detector.
Since normally no colluder is willing to take more of a risk than any other colluder,
averaging attack in which ${\lambda}_{j_{\ell}} = 1/t$, $1 \leq {\ell} \leq t$, is the most fair choice
for each colluder to avoid detection, as claimed in \cite{LTWWZ,TWWL}.
This attack also makes the pirate copy have better perceptional quality.
Based on the discussions above, the observed content ${\bf y}$ after averaging attack  is
\[ {\bf y} = \frac{1}{t}\sum\limits_{{\ell}=1}^{t}{\bf y}_{j_{\ell}} = \frac{\alpha}{t}\sum\limits_{{\ell}=1}^{t}{\bf w}_{j_{\ell}} + {\bf x}=
 \alpha\sum\limits_{{\ell}=1}^{t}\sum\limits_{i=1}^{n}\frac{b_{ij_{\ell}}}{t}{\bf u}_{i} + {\bf x}.\]

%Due to the orthogonality of the orthonormal basis $\{{\bf u}_i \ | \ 1 \leq i \leq n\}$,
In colluder detection phase,
we compute the correlation vector
${\bf T} = ( {\bf T}(1), {\bf T}(2)$, $\ldots, {\bf T}(n))$,
where ${\bf T}(i) = \langle \frac{{\bf y}-{\bf x}}{\alpha}$, ${\bf u}_{i}\rangle$, $1 \leq i \leq n$,
and $\langle \frac{{\bf y}-{\bf x}}{\alpha}, {\bf u}_{i}\rangle$ is the inner product of $\frac{{\bf y}-{\bf x}}{\alpha}$ and  ${\bf u}_i$.
We would like to strategically design an anti-collusion code  to accurately identify the contributing fingerprints
involved in the averaging attack from this detection statistics ${\bf T}$.
Note that in code design phase, we only need to consider deterministic anti-collusion codes over a finite set.

%%%%%%%%%%%%%%%%%%%%%%%%%%%%%%%%%%%%%%%%%%%
\section{Frameproof Codes and  Strongly Separable Codes}   %
%%%%%%%%%%%%%%%%%%%%%%%%%%%%%%%%%%%%%%%%%%%
\label{FPCSSC}

In this section, we first recall the traceability of frameproof codes ($t$-FPCs) for multimedia contents.
Then we introduce the notion of a strongly separable code ($\overline{t}$-SSC).
We show that a binary $\overline{t}$-SSC
has weaker requirements than a binary $t$-FPC
but has the same traceability as a binary $t$-FPC.
We also present a composition construction for  binary $\overline{t}$-SSCs
from $q$-ary $\overline{t}$-SSCs.

Let $n, M$ and $q$ be positive integers, and $Q$ an alphabet with $|Q|=q$.
A set $\mathcal{C} = \{{\bf c}_1,{\bf c}_2,\ldots, {\bf c}_M\} \subseteq Q^n$ is called an $(n,M,q)$ code
and each ${\bf c}_i$ is called a codeword.
Without loss of generality, we may assume $Q=\{0,1,\ldots,q-1\}$.
When $Q=\{0,1\}$, we also use the word ``binary".
Given an $(n,M,q)$ code, its incidence matrix is the $n \times M$ matrix on $Q$
in which the columns are the $M$ codewords in $\mathcal{C}$.
Often, we make no difference between an $(n,M,q)$ code and its incidence matrix unless otherwise stated.

For any code $\mathcal{C} \subseteq Q^n$, we define the set of $i$th coordinates of $\mathcal{C}$ as
\[ \mathcal{C}(i) =\{{\bf c}(i) \in Q \ | \ {\bf c}=({\bf c}(1), {\bf c}(2), \ldots, {\bf c}(n))^{T} \in \mathcal{C}\}\]
for any $1 \le i \le n$.
For any subset of codewords $\mathcal{C}^{'} \subseteq \mathcal{C}$, we define the descendant code of $\mathcal{C}^{'}$ as
\begin{eqnarray*}
 {\sf desc}(\mathcal{C}^{'}) = \{({\bf x}(1), {\bf x}(2), \ldots, {\bf x}(n) )^{T}  \in Q^n \ |   \  {\bf x}(i) \in \mathcal{C}^{'}(i), 1 \le i \le n\},\end{eqnarray*}
that is,
\[ {\sf desc}(\mathcal{C}^{'})=\mathcal{C}^{'}(1) \times \mathcal{C}^{'}(2) \times \cdots \times  \mathcal{C}^{'}(n).\]
The set ${\sf desc}(\mathcal{C}^{'})$ consists of the $n$-tuples that could be
produced by a coalition holding the codewords in $\mathcal{C}^{'}$.

Using the notions of descendant codes and sets of $i$th coordinates of codes,
we can define frameproof codes which were first introduced in \cite{BS}.

\begin{definition} \rm
\label{defFPC}
Let $\mathcal{C}$ be an $(n,M,q)$ code and $t \ge 2$ be an integer.
$\mathcal{C}$ is a $t$-frameproof code, or $t$-FPC$(n,M,q)$, if
for any $\mathcal{C}' \subseteq \mathcal{C}$ such that $|\mathcal{C}'| \le t$,
we have ${\sf desc}(\mathcal{C}') \bigcap \mathcal{C} = \mathcal{C}'$,
that is, for any ${\bf c} =({\bf c}(1),\ldots,{\bf c}(n))^{T} \in \mathcal{C} \setminus \mathcal{C}'$,
there is at least one coordinate $i$, $1 \le i \le n$,
such that ${\bf c}(i) \not\in  \mathcal{C}'(i)$.
\end{definition}

Intuitively, an $(n,M,q)$ code is a $t$-FPC if no coalition of size at most $t$ can frame another user not in
the coalition by producing the codeword held by that user in generic digital fingerprinting. Frameproof codes were widely considered as
having no traceability for generic digital data (see for example \cite{STW}). However, Cheng and Miao \cite{CM}
showed that frameproof codes actually have traceability for multimedia contents.
The main reason is explained below.

In the multimedia scenario,
for any set of colluders holding codewords $\mathcal{C}_0 \subseteq \mathcal{C}$ and for any index $1 \leq i \leq n$,
their detection statistics ${\bf T}(i)$ mentioned in Section \ref{pre} possesses the whole information
on $\mathcal{C}_0(i)$; namely, we have ${\bf T}(i)=1$ if and only if $\mathcal{C}_0(i)=\{1\}$,
${\bf T}(i)=0$ if and only if $\mathcal{C}_0(i)=\{0\}$, and
$0 < {\bf T}(i) < 1$ if and only if $\mathcal{C}_0(i)=\{0,1\}$.
Therefore,
%we also call $R = \mathcal{C}_0(1) \times \cdots \times \mathcal{C}_0(n)
% \subseteq \mathcal{C}(1) \times \cdots \times \mathcal{C}(n)$
%the descendant code derived from the detection statistics ${\bf T}$.
%In a word,
we can capture a set $R = \mathcal{C}_0(1) \times \cdots \times \mathcal{C}_0(n)
 \subseteq \mathcal{C}(1) \times \cdots \times \mathcal{C}(n)$ in the multimedia
scenario from the detection statistics ${\bf T}$, instead of an element ${\bf d} \in R$ in the generic digital scenario.
The property a frameproof code holds makes it easy to identify $\mathcal{C}_0$, and thus the set of colluders holding $\mathcal{C}_0$
who have produced $R$.

\begin{lemma} $(${\rm \cite{CM}}$)$
\label{algFPC}
Under the assumption that the number of colluders in the averaging attack is at most $t$,
any $t$-{\rm FPC}$(n, M, 2)$  can be used to identify all the colluders with computational
complexity $O(nM)$ by using the algorithm {\tt LACCIdenAlg}  in {\rm \cite{CM}}.
\end{lemma}
%%%%%%%%%%%%%%%    single column   %%%%%%%%%%%%%%%%%%%%%%%%%%%%%%%%%%%%%
\begin{algorithm}%[H]
\caption{ {\tt LACCIdenAlg}(${R}$)}
\label{AlgFPC}
Define $J_a$, $J_o$ to be the sets of indices where $R(j) = \{1\}$, $R(j) = \{0\}$, respectively,
and  ${\bf J_a} = ({\bf J_a}(1), \ldots, {\bf J_a}(|J_a|))^T$,  ${\bf J_o} = ({\bf J_o}(1), \ldots, {\bf J_o}(|J_o|))^T$
to be the vector representing $R$'s coordinates where $R(j) = \{1\}$ and $R(j) = \{0\}$, respectively\;
${\bf \Phi } = {\bf 1}$\;
$U_1 = \emptyset$\;
\For { $k=1$  {\bf to }  $|J_a|$ }
     {   $j = {\bf J_a}(k)$\;
         define ${\bf e}_j$ to be the $j$th row of $\mathcal{C}$\;
         ${\bf \Phi } = {\bf \Phi} \cdot {\bf e}_j$\;
     }

\For { $i=1$ {\bf to} $M$}
     {   \If {${\bf \Phi}(i) = 1$}
         {$U_1 = \{i\} \bigcup U_1$\;}
     }

${\bf \Phi } = {\bf 1}$\;
$U_2 = \emptyset$\;

\For { $k=1$ {\bf to} $|J_o|$}
     {   $j = {\bf J_o}(k)$\;
         ${\bf \Phi } = {\bf \Phi } \cdot \overline{{\bf e}}_j$\;
     }

\For { $i=1$ {\bf to} $M$}
     {   \If {${\bf \Phi}(i) = 1$}
             {$U_2 = \{i\} \bigcup U_2$\;
             }
     }

$U = U_1 \bigcap U_2$\;

\eIf { $|U| \leq t$ }
    { {\bf output }$U$\;}
    { {\bf output} ``The set of colluders has size at least $t+1$."}
\end{algorithm}
The multimedia fingerprinting scheme based on a $t$-FPC$(n, M, 2)$ can have at most
$r\cdot2^{\lceil\frac{n}{t}\rceil} + O(2^{\lceil\frac{n}{t}\rceil-1})$ authorized users,
where $r$ is the unique integer such that $r \in \{0, 1, \ldots, t-1\}$ and $r \equiv n \pmod t$ \cite{Bla}.
In most cases, this number of users is still too small to be of practical use.

We now pay our attention to a new concept of a strongly separable code defined below, which can support
a multimedia fingerprinting scheme with more users than a $t$-FPC does.

\begin{definition}
\label{SSC}
Let $\mathcal{C}$ be an $(n, M, q)$ code and $t \ge 2$ be an integer.
$\mathcal{C}$ is a strongly $\overline{t}$-separable code, or $\overline{t}$-{\rm SSC}$(n,M,q)$, if for any
$\mathcal{C}_0 \subseteq \mathcal{C}$, $1 \le |\mathcal{C}_0| \leq t$,
we have $\bigcap_{\mathcal{C}^{'} \in S(\mathcal{C}_0)}\mathcal{C}^{'} = \mathcal{C}_0$,
where $S(\mathcal{C}_0) = \{ \mathcal{C}^{'} \subseteq \mathcal{C} \ | \ {\rm desc}(\mathcal{C}^{'}) = {\rm desc}(\mathcal{C}_0) \}$.
\end{definition}

From the definition, it is clear that for any $\mathcal{C}^{'} \in S(\mathcal{C}_0)$ and $\mathcal{C}^{'} \neq \mathcal{C}_0$,
we have $\mathcal{C}_0 \subseteq \mathcal{C}^{'}$ and $|\mathcal{C}^{'}| \ge t+1$.

The following theorem shows that a binary $\overline{t}$-SSC can be used to identify all the colluders
in the averaging attack with computational complexity $O(nM)$, which is the same as that of a binary $t$-FPC.

\begin{theorem}
\label{algSSC}
Under the assumption that the number of colluders in the averaging attack is at most $t$,
any  $\overline{t}$-{\rm SSC}$(n, M, 2)$ can be used to identify all the colluders  with computational complexity $O(nM)$
by applying Algorithm {\rm \ref{AlgSSC}} described below.
\end{theorem}
\proof Let $\mathcal{C}$ be the  $\overline{t}$-SSC$(n, M, 2)$, and $R$ be the descendant code derived
from the detection statistics ${\bf T}$.
Then by applying the following tracing algorithm, Algorithm \ref{AlgSSC},
%for any given outcome binary OR vector ${\bf r}_{\rm{OR}}$ and AND vector ${\bf r}_{\rm{AND}}$,
one can identify all the colluders. The computational complexity is clearly  $O(nM)$.
\begin{algorithm}%[H]
\caption{ {\tt SSCTraceAlg}(${R}$)}
\label{AlgSSC}
Define $J_a$, $J_o$ to be the sets of indices where $R(j) = \{1\}$, $R(j) = \{0\}$, respectively,
and  ${\bf J_a} = ({\bf J_a}(1), \ldots, {\bf J_a}(|J_a|))^T$,  ${\bf J_o} = ({\bf J_o}(1), \ldots, {\bf J_o}(|J_o|))^T$
to be the vector representing $R$'s coordinates where $R(j) = \{1\}$ and $R(j) = \{0\}$, respectively\;
${\bf \Phi } = {\bf 1}$\;
$U_a = \emptyset$\;
$U_o = \emptyset$\;
$U = \emptyset$\;
\For { $k=1$  {\bf to }  $|J_a|$ }
     {   $j = {\bf J_a}(k)$\;
         define ${\bf e}_j$ to be the $j$th row of $\mathcal{C}$\;
         ${\bf \Phi } = {\bf \Phi} \cdot {\bf e}_j$\;
     }

\For { $k=1$ {\bf to} $|J_o|$}
     {   $j = {\bf J_o}(k)$\;
         ${\bf \Phi } = {\bf \Phi } \cdot \overline{{\bf e}}_j$\;
     }

\For { $k = 1$ {\bf to} $n$ }
     {   ${\bf \Phi}_a = {\bf \Phi} \cdot {\bf e}_k$\;
         ${\bf \Phi}_o = {\bf \Phi} \cdot \overline{{\bf e}}_k$\;
         \For { $i = 1$ {\bf to} $M$ }
              {   \If { ${\bf \Phi}_a(i) = 1$ }
                      {   $U_a = \{i\} \bigcup U_a$\;
                      }
              }
         \If  { $|U_a|=1$ }
              {   $U = U \bigcup U_a$\;
              }
         \For { $i = 1$ {\bf to} $M$ }
              {   \If { ${\bf \Phi}_o(i) = 1$}
                      {   $U_o = \{i\} \bigcup U_o$\;
                      }
              }

         \If { $|U_o|=1$  }
             { $U = U \bigcup U_o$\;
             }
    }

\eIf { $|U| \leq t$ }
    { {\bf output }$U$\;}
    { {\bf output} ``The set of colluders has size at least $t+1$."}
\end{algorithm}

According to  Algorithm \ref{AlgSSC}, by deleting all columns
$\{ {\bf c} \in \mathcal{C} \  | \    \exists  \ 1 \leq i \leq n,
R(i) = \{1\}, {\bf c}(i) = 0, \ {\rm or} \ R(i) = \{0\}, {\bf c}(i) = 1\}$, we obtain a sub-matrix $\mathcal{C}_{L}$ of $\mathcal{C}$.
Suppose that $C_0 = \{u_1, u_2, \ldots, u_r\}$, $1 \leq r \leq t$,
is the set of colluders, the codeword ${\bf c}_i$ is assigned to the colluder $u_i$,  $1 \leq i \leq r$,
and $\mathcal{C}_0 = \{ {\bf c}_1, {\bf c}_2, \ldots, {\bf c}_r\}$.
It is not difficult to see that $\mathcal{C}_0 \subseteq \mathcal{C}_{L}$.
According to the definition of a $\overline{t}$-SSC,
we have $\bigcap_{\mathcal{C}^{'} \in S(\mathcal{C}_0 )}\mathcal{C}^{'} = \mathcal{C}_0 \neq \emptyset$,
where $S(\mathcal{C}_0) = \{ \mathcal{C}^{'} \subseteq \mathcal{C} \ | \ {\sf desc}(\mathcal{C}^{'}) = {\sf desc}(\mathcal{C}_0) = R \}$.
We prove this theorem in three steps.

(1)   $\mathcal{C}_{L} \in S(\mathcal{C}_0)$, that is, ${\sf desc}(\mathcal{C}_{L}) = R$.
For any $1 \leq j \leq n$, we consider the following cases.

(1.1)   $R(j) = \{1\}$. For any
${\bf c} \in \mathcal{C}_{L} $, ${\bf c}(j)=1$ according to the processes deriving $\mathcal{C}_{L}$.
So $\mathcal{C}_{L}(j) = \{1\} = R(j)$.

(1.2)   $R(j) = \{0\}$. For any ${\bf c} \in \mathcal{C}_{L}$, ${\bf c}(j)=0$
according to the processes  deriving $\mathcal{C}_{L}$.
So $\mathcal{C}_{L}(j) = \{0\} = R(j)$.

(1.3)  $R(j) = \{0, 1\}$.  Since ${\sf desc}(\mathcal{C}_{0}) = R$,
we know that there exist ${\bf c}_1, {\bf c}_2 \in \mathcal{C}_0 \subseteq \mathcal{C}_{L}$
such that ${\bf c}_1(j) =0$ and ${\bf c}_2(j) =1$, respectively.
This implies $\mathcal{C}_{L}(j) = \{0, 1\} = R(j)$.

According to (1.1)-(1.3) above, for any $1 \leq j \leq n$, we have
$\mathcal{C}_{L}(j)= R(j)$,
which implies ${\sf desc}(\mathcal{C}_{L}) = R$.

(2)   We want to  show that  for any
${\bf x} \in \mathcal{C}_0 =\bigcap_{\mathcal{C}^{'} \in S(\mathcal{C}_0 )}\mathcal{C}^{'}$,
there exists $1 \leq j \leq n$, such that  ${\bf x}(j) =1$ and ${\bf c}(j) = 0$
for any ${\bf c} \in \mathcal{C}_{L} \setminus \{ {\bf x}\}$,
or ${\bf x}(j) =0$ and ${\bf c}(j) = 1$ for any ${\bf c} \in \mathcal{C}_{L} \setminus \{ {\bf x}\}$.
Assume not. Then for any  $1 \leq j \leq n$,  ${\bf x}(j) =1$ implies that there exists
${\bf c}_1 \in \mathcal{C}_{L} \setminus \{ {\bf x} \}$ such that ${\bf c}_1(j) = 1$,
and ${\bf x}(j) =0$ implies that there exists
${\bf c}_2\in \mathcal{C}_{L} \setminus \{ {\bf x} \}$ such that ${\bf c}_2(j) = 0$.
Then we have  ${\sf desc}(\mathcal{C}_{L}) = {\sf desc}(\mathcal{C}_{L}\setminus\{ {\bf x} \})$.
Since $\mathcal{C}_{L} \in S(\mathcal{C}_0)$ by (1),
we can have  $\mathcal{C}_{L} \setminus \{ {\bf x}\} \in S(\mathcal{C}_0)$, and
${\bf x} \notin \bigcap_{{ \mathcal{C}}^{'} \in S(\mathcal{C}_0)} \mathcal{C}^{'}$, a contradiction.

(3)   At last, according to Algorithm \ref{AlgSSC} and (2), it suffices to show that any user $u$
assigned with a codeword ${\bf x} \in \mathcal{C}_0 = \bigcap_{\mathcal{C}^{'} \in S(\mathcal{C}_0)}\mathcal{C}^{'}$
is a colluder. Assume that $u$ is not a colluder.
Then for any $\mathcal{C}^{'} \in S(\mathcal{C}_0)$,
we have $\mathcal{C}^{'} \setminus\{{\bf x}\} \in S(\mathcal{C}_0)$,
which implies ${\bf x} \notin \bigcap_{\mathcal{C}^{'}  \in S(\mathcal{C}_0)}\mathcal{C}^{'}$,
a contradiction.

This completes the proof.
\qed

We now consider the relationship between a $t$-FPC$(n, M, q)$ and a $\overline{t}$-{\rm SSC}$(n, M, q)$.

\begin{lemma}
\label{rela2}
Any $t$-{\rm FPC}$(n, M, q)$ is a $\overline{t}$-{\rm SSC}$(n, M, q)$.
\end{lemma}
\proof  Let  $\mathcal{C}$ be a $t$-FPC$(n, M, q)$.
We are going to show that for any
$\mathcal{C}_0 \subseteq \mathcal{C}$, $|\mathcal{C}_0| \leq t$,
$S(\mathcal{C}_0) = \{ \mathcal{C}^{'} \subseteq \mathcal{C} \ | \ {\sf desc}(\mathcal{C}^{'})
= {\sf desc}(\mathcal{C}_0) \} = \{\mathcal{C}_0\}$.
Assume that there exists $\mathcal{C}^{'} \in S(\mathcal{C}_0)$ such that $\mathcal{C}^{'} \neq \mathcal{C}_0$.

(1) \ If $|\mathcal{C}^{'}| \geq |\mathcal{C}_0|$,
then there exists ${\bf c} \in \mathcal{C}^{'} \subseteq \mathcal{C}$ such that ${\bf c} \notin \mathcal{C}_0$.
Since ${\sf desc}(\mathcal{C}^{'}) = {\sf desc}(\mathcal{C}_0)$, we have ${\bf c} \in {\sf desc}(\mathcal{C}_0) \bigcap \mathcal{C}$,
while  $|\mathcal{C}_0| \leq t$, a contradiction to the definition of a $t$-{\rm FPC}.

(2) \ If $|\mathcal{C}^{'}| < |\mathcal{C}_0| \leq t$, then there exists ${\bf c} \in \mathcal{C}_0 \subseteq \mathcal{C}$
such that ${\bf c} \notin \mathcal{C}^{'}$.
Since ${\sf desc}(\mathcal{C}^{'}) = {\sf desc}(\mathcal{C}_0)$, we have ${\bf c} \in {\sf desc}(\mathcal{C}^{'}) \bigcap \mathcal{C}$,
while  $|\mathcal{C}^{'}| < t$, a contradiction to the definition of a $t$-{\rm FPC}.

According to the discussions above, we have $S(\mathcal{C}_0) = \{\mathcal{C}_0\}$,
which implies $\bigcap_{\mathcal{C}^{'} \in S(\mathcal{C}_0)}\mathcal{C}^{'} = \mathcal{C}_0$.
\qed

The following example shows that the converse of Lemma \ref{rela2} does not always hold.

\begin{exmp} Consider the following $(3, 4, 2)$ code $\mathcal{C}$:
\begin{eqnarray*}
\begin{array}{c}
  \ \ \ \ \ {\bf c}_1 \ \ {\bf c}_2\ \  \ {\bf c}_3   \ \ {\bf c}_4 \\
\mathcal{C}=
\left(\begin{array}{cccc}
 0 \ & 1 \ & 0 \ & 0 \ \\
 0 \ & 0 \ & 1 \ & 0  \ \\
 0 \ & 0 \ & 0 \ & 1  \
  \end{array}\right)
  \end{array}
\end{eqnarray*}
We can directly check that $\mathcal{C}$ is a $\overline{2}$-{\rm SSC}$(3, 4, 2)$.
Now, we show that $\mathcal{C}$ is not a $2$-{\rm FPC}.
For $\mathcal{C}^{'} = \{{\bf c}_2, {\bf c}_3\}$,
${\sf desc}(\mathcal{C}') \bigcap \mathcal{C} = \{{\bf c}_1, {\bf c}_2, {\bf c}_3\} \neq \mathcal{C}'$.
This is a contradiction to the definition of a $2$-{\rm FPC}.
\end{exmp}

We would like to make a remark here. From Theorem \ref{algSSC}, Lemmas \ref{algFPC}, \ref{rela2},
we can know that $\overline{t}$-{\rm SSC}$(n, M, 2)$s have weaker requirements than $t$-{\rm FPC}$(n, M, 2)$s but have the
same traceability as $t$-{\rm FPC}$(n, M, 2)$s.

At the end of this section, we show a composition construction for binary $\overline{t}$-SSCs from $q$-ary $\overline{t}$-SSCs,
which makes the study of $q$-ary $\overline{t}$-SSCs with short length, say $n=2,3$, important.

\begin{lemma}
\label{compconstru}
If there exists a $\overline{t}$-{\rm SSC}$(n, M, q)$, then there exists a $\overline{t}$-{\rm SSC}$(nq, M, 2)$.
\end{lemma}
\proof  Let $\mathcal{C} = \{{\bf c}_1, {\bf c}_2, \ldots, {\bf c}_M\}$
be a $\overline{t}$-SSC$(n, M, q)$ on $Q = \{ 0, 1, \ldots, q - 1\}$, and
$E = \{{\bf e}_1, {\bf e}_2, \ldots, {\bf e}_{q}\}$, where ${\bf e}_i$ is the $i$th identity vector of length $q$. Let $f$: $Q \rightarrow E$
 be a bijective mapping such that $f(i) = {\bf e}_{i+1}$.
For any codeword ${\bf c} = ({\bf c}(1), {\bf c}(2), \ldots, {\bf c}(n))^{T} \in \mathcal{C}$, we define
$f({\bf c}) = (f({\bf c}(1)), f({\bf c}(2)), \ldots, f({\bf c}(n)))^{T}$.
Obviously, $f({\bf c})$ is a binary vector of length $nq$. We define a new $(nq, M, 2)$ code
$\mathcal{F} = \{ f({\bf c}_1), f({\bf c}_2), \ldots, f({\bf c}_M)\}$.
We can show that $\mathcal{F}$ is a $\overline{t}$-SSC.

For any $\mathcal{F}_0 \subseteq \mathcal{F}$, $|\mathcal{F}_0| \leq t$, we only need to show that for any
$\mathcal{F}_1 \subseteq \mathcal{F}$, ${\sf desc}(\mathcal{F}_0) = {\sf desc}(\mathcal{F}_1)$ implies
$\mathcal{F}_0 \subseteq \mathcal{F}_1$. Suppose $\mathcal{F}_0$, $\mathcal{F}_1$ correspond to two codeword
sets $\mathcal{C}_0, \mathcal{C}_1 \subseteq \mathcal{C}$, respectively, such that $|\mathcal{C}_0| =  |\mathcal{F}_0| \leq t$,
where $\mathcal{F}_0 = \{ f({\bf c}) \ | \ {\bf c} \in \mathcal{C}_0\}$ and $\mathcal{F}_1 = \{ f({\bf c}) \ | \ {\bf c} \in \mathcal{C}_1\}$.
Since desc$(\mathcal{F}_0) = $ desc$(\mathcal{F}_1)$, we have desc$(\mathcal{C}_0) = $ desc$(\mathcal{C}_1)$.
Then $\mathcal{C}_0 \subseteq \mathcal{C}_1$, because ${ \mathcal{C}}$ is a $\overline{t}$-SSC$(n, M, q)$.
So, $\mathcal{F}_0 \subseteq \mathcal{F}_1$.

This completes the proof.
\qed

%%%%%%%%%%%%%%%%%%%%%%%%%%%%%%%%%%%%%%%%%%%
\section{Constructions for Strongly Separable Codes}   %
%%%%%%%%%%%%%%%%%%%%%%%%%%%%%%%%%%%%%%%%%%%
\label{constr}

Let $M(\overline{t},n,q)$  denote the maximum number of codewords
in a $\overline{t}$-SSC$(n, M, q)$. A $\overline{t}$-SSC$(n,M,q)$ is optimal if $M = M(\overline{t},n,q)$.
In this section, the relationship between strongly separable codes and separable codes is considered, and
several infinite series of optimal $\overline{2}$-SSCs of length $2$ are derived based on this relationship.
We also obtain the forbidden configurations of $\overline{2}$-{\rm SSC}$(3, M, q)$s.
Finally, we present a construction for $\overline{2}$-{\rm SSC}$(3, M, q)$s, where $M$ is nearly equal to $\frac{9}{8}q^2$.

\subsection{Relationship between Strongly Separable Codes and Separable Codes}

In this subsection, we first recall the concept of a separable code introduced in \cite{CM}
and investigated in detail in \cite{CJM, GG},
which is a powerful tool to investigate SSCs.

\begin{definition} $(${\rm \cite{CM}}$)$
\label{defSC}
Let $\mathcal{C}$ be an $(n,M,q)$ code and $t \ge 2$ be an integer.
$\mathcal{C}$ is a $\overline{t}$-separable code, or $\overline{t}$-{\rm SC}$(n,M,q)$,
if for any $\mathcal{C}_1, \mathcal{C}_2 \subseteq \mathcal{C}$ such that
$1 \le |\mathcal{C}_1| \le t$, $1 \le |\mathcal{C}_2| \le t$ and $\mathcal{C}_1 \neq \mathcal{C}_2$,
we have ${\sf desc}(\mathcal{C}_1) \neq {\sf desc}(\mathcal{C}_2)$,
that is, there is at least one coordinate $i$, $1 \le i \le n$, such that $\mathcal{C}_1(i) \neq \mathcal{C}_2(i)$.
\end{definition}

Recall that for any $\mathcal{C}^{'} \in S(\mathcal{C}_0)$ and $\mathcal{C}^{'} \neq \mathcal{C}_0$,
we have $\mathcal{C}_0 \subseteq \mathcal{C}^{'}$ and $|\mathcal{C}^{'}| \ge t+1$.
In other words, for any $\overline{t}$-SSC$(n,M,q)$, $\mathcal{C}$, there are no distinct subsets
$\mathcal{C}_1, \mathcal{C}_2 \subseteq \mathcal{C}$ with  $1 \le |\mathcal{C}_1| \le t$, $1 \le |\mathcal{C}_2| \le t$, such that
${\sf desc}(\mathcal{C}_1) = {\sf desc}(\mathcal{C}_2)$. This implies  the following lemma.

\begin{lemma}
\label{rela1}
Any $\overline{t}$-{\rm SSC}$(n, M, q)$ is a $\overline{t}$-{\rm SC}$(n, M, q)$.
\end{lemma}

The following example shows that the converse of Lemma \ref{rela1} does not always hold.

\begin{exmp} Consider the following $(3, 5, 2)$ code $\mathcal{C}$:
\begin{eqnarray*}
\begin{array}{c}
  \ \ \ \ {\bf c}_1 \ \ {\bf c}_2\ \  \ {\bf c}_3   \ \ {\bf c}_4 \ \ \ {\bf c}_5 \\
\mathcal{C}=
\left(\begin{array}{ccccc}
 0 \ & 1 \ & 0 \ & 0 \ & 1 \ \\
 0 \ & 0 \ & 1 \ & 0 \ & 1 \ \\
 0 \ & 0 \ & 0 \ & 1 \ & 1 \
  \end{array}\right)
  \end{array}
\end{eqnarray*}
We can directly check that $\mathcal{C}$ is a $\overline{2}$-{\rm SC}$(3, 5, 2)$.
Now, we show that $\mathcal{C}$ is not a $\overline{2}$-{\rm SSC}.
Let $\mathcal{C}_0 = \{{\bf c}_1, {\bf c}_5\}$ and $\mathcal{C}^{'}= \{{\bf c}_2, {\bf c}_3, {\bf c}_4\}$,
then ${\sf desc}(\mathcal{C}_0) = {\sf desc}(\mathcal{C}^{'})$, while $\mathcal{C}_0 \not\subseteq  \mathcal{C}^{'}$.
This implies that $\mathcal{C}$ is not a $\overline{2}$-{\rm SSC}$(3, 5, 2)$.
\end{exmp}

%\indent\indent\indent\indent \indent \indent\indent ${\bf c}_1  \  \ {\bf c}_2 \  {\bf c}_3 \ \ {\bf c}_4 \  {\bf c}_5 $
%$$\mathcal{C}=
%\left( \begin{array}{ccccc}
%0 & 1 &  0 &  0  &  1\\
%0 & 0 &  1 &  0  &  1\\
%0 & 0 &  0 &  1  &  1
%\end{array} \right)$$

However, the following result shows that a $\overline{2}$-SSC$(2,M,q)$ is always a $\overline{2}$-SC$(2,M,q)$.

\begin{theorem}
\label{equi4}
Let $\mathcal{C}$ be a $(2,M,q)$ code.
Then $\mathcal{C}$ is a $\overline{2}$-{\rm SSC}$(2,M,q)$ if and only if it is a $\overline{2}$-{\rm SC}$(2,M,q)$.
\end{theorem}
\proof By Lemma \ref{rela1}, it suffices to consider the sufficiency.
Let $\mathcal{C}$ be a $\overline{2}$-SC$(2, M, q)$.
Assume that $\mathcal{C}$ is not a $\overline{2}$-SSC$(2, M, q)$. Then
there exist $\mathcal{C}_0, \mathcal{C}^{'} \subseteq \mathcal{C}$, $|\mathcal{C}_0| \leq 2$, such that
${\sf desc}(\mathcal{C}_0) = {\sf desc}(\mathcal{C}^{'})$ but $\mathcal{C}_0 \not\subseteq \mathcal{C}^{'}$.
If $|\mathcal{C}_0|=1$, then it is clear that $\mathcal{C}_0 = \mathcal{C}^{'}$, a contradiction.
So $|\mathcal{C}_0|=2$. Let $\mathcal{C}_0 = \{ {\bf c}_1, {\bf c}_2 \}$, ${\bf c}_i = (a_i, b_i)^{T}$, where $i = 1, 2$.
Since $\mathcal{C}$ is a $\overline{2}$-{\rm SC}$(2, M, q)$ and
${\sf desc}(\mathcal{C}_0) = {\sf desc}(\mathcal{C}^{'})$,
we  have $\mathcal{C}^{'} \subseteq {\sf desc}(\mathcal{C}_0) \bigcap \mathcal{C}$
and $|\mathcal{C}^{'}| \geq 3$. We  now consider the Hamming distance $d({\bf c}_1, {\bf c}_2)$ of ${\bf c}_1$ and ${\bf c}_2$  .

(1)  If $d({\bf c}_1, {\bf c}_2) = 1$, without loss of generality, we may assume $a_1 = a_2$, $b_1 \neq b_2$. Then $|{\sf desc}(\mathcal{C}_0)| = 2$.
So $ |\mathcal{C}^{'}| \leq  |{\sf desc}(\mathcal{C}_0)| =2$, a contradiction.

(2)  If $d({\bf c}_1, {\bf c}_2) = 2$, then $a_1 \neq a_2$, $b_1 \neq b_2$,
and ${\sf desc}(\mathcal{C}_0) = \{ {\bf c}_1, {\bf c}_2, {\bf c}_3, {\bf c}_4 \}$,
where ${\bf c}_3 = (a_1, b_2)^{T}$ and ${\bf c}_4 = (a_2, b_1)^{T}$.
Then $|{\sf desc}(\mathcal{C}_0) \bigcap \mathcal{C}| \leq 3$.
Otherwise, if $|{\sf desc}(\mathcal{C}_0) \bigcap \mathcal{C}| = 4$, i.e.,
${\sf desc}(\mathcal{C}_0) \bigcap \mathcal{C} = \{ {\bf c}_1, {\bf c}_2, {\bf c}_3, {\bf c}_4 \}$,
then ${\sf desc}(\{ {\bf c}_1,$ $ {\bf c}_2\})= {\sf desc}(\{ {\bf c}_3, {\bf c}_4\})$,
a contradiction to the definition of a $\overline{2}$-{\rm SC}.
Since $\mathcal{C}^{'} \subseteq {\sf desc}(\mathcal{C}_0) \bigcap \mathcal{C}$ and $|\mathcal{C}^{'}| \geq 3$,
we have $|\mathcal{C}^{'}| = 3$. So, we may assume, without loss of generality,
that $\mathcal{C}^{'} =  \{ {\bf c}_1, {\bf c}_2, {\bf c}_3\}$,
which implies $\mathcal{C}_0 \subseteq \mathcal{C}^{'}$, a contradiction.

This completes the proof.
\qed

$\overline{2}$-SCs were well studied in \cite{CFJLM1, CJM},
and several infinite series of optimal $\overline{2}$-SC$(2,M,q)$s were constructed.

\begin{lemma} $(${\rm \cite{CFJLM1, CJM}}$)$
\label{reulSC}
Let $k \geq 2$ be a prime power.
Then there is an optimal  $\overline{2}$-{\rm SC}$(2,M,q)$ for any $q \in \{ k^2 -1, k^2+k-2, k^2+k-1, k^2+k, k^2+k+1\}$.
\end{lemma}

These optimal SCs are, in fact, optimal SSCs from the equivalence stated in Theorem \ref{equi4}.

\begin{corollary}
\label{optiSSC}
Let $k \geq 2$ be a prime power.
Then there is an optimal  $\overline{2}$-{\rm SSC}$(2,M,q)$ for any $q \in \{ k^2 -1, k^2+k-2, k^2+k-1, k^2+k, k^2+k+1\}$.
\end{corollary}

We note that a $2$-FPC$(2, M, q)$ can roughly have at most $2q$ codewords \cite{Bla}, but the
optimal $\overline{2}$-SSC$(2,M,q)$s in Corollary \ref{optiSSC} can have about $q^\frac{3}{2}$ codewords \cite{CFJLM1, CJM}.

%%%%%%%%%%%%%%%%%%%%%%%%%%%%
\subsection{Constructions for $\overline{2}$-SSC$(3, M, q)$s} %
%%%%%%%%%%%%%%%%%%%%%%%%%%%%

From Lemma \ref{rela1}, any $\overline{2}$-SSC$(3,$ $  M, q)$ is a $\overline{2}$-SC$(3, M,q)$. Therefore,
we can start from $\overline{2}$-SC$(3, M, q)$s to investigate $\overline{2}$-SSC$(3, M, q)$s.
At first, we derive forbidden configurations of a $\overline{2}$-SSC$(3, M, q)$.

\begin{lemma}
\label{forbiconfi1}
Let $\mathcal{C}$ be a $\overline{2}$-{\rm SC}$(3, M, q)$. If there exist $\mathcal{C}_0, \mathcal{C}^{'} \subseteq \mathcal{C}$,
$|\mathcal{C}_0| \leq 2$ such that ${\sf desc}(\mathcal{C}_0) = {\sf desc}(\mathcal{C}^{'})$ and
$\mathcal{C}_0 \not\subseteq  \mathcal{C}^{'}$, then  $\mathcal{C}_0 = \{ {\bf c}_1, {\bf c}_2 \}$ and
the Hamming distance $d({\bf c}_1, {\bf c}_2) \notin \{0, 1, 2\}$.
\end{lemma}
\proof If $|\mathcal{C}_0|=1$, then it is clear that $\mathcal{C}_0 = \mathcal{C}^{'}$, a contradiction.
So $|\mathcal{C}_0|=2$. Let $\mathcal{C}_0 = \{ {\bf c}_1, {\bf c}_2 \}$, ${\bf c}_i = (a_i, b_i, e_i)$, ${\bf c}_1 \ne {\bf c_2}$.
Since $\mathcal{C}$ is a $\overline{2}$-{\rm SC}$(3, M, q)$ and ${\sf desc}(\mathcal{C}_0) = {\sf desc}(\mathcal{C}^{'})$,
we  have $\mathcal{C}^{'} \subseteq {\sf desc}(\mathcal{C}_0) \bigcap \mathcal{C}$ and $|\mathcal{C}^{'}| \geq 3$.

(1)   If $d({\bf c}_1, {\bf c}_2) = 1$, we may assume, without loss of generality,
that $a_1 = a_2$, $b_1 = b_2$, $e_1 \neq e_2$. Then $|{\sf desc}(\mathcal{C}_0)| = 2$.
So $ |\mathcal{C}^{'}| \leq  |{\sf desc}(\mathcal{C}_0)| =2$, a contradiction.

(2)   If $d({\bf c}_1, {\bf c}_2) = 2$, we may assume, without loss of generality,  $a_1 = a_2$, $b_1 \neq b_2$, $e_1 \neq e_2$.
Then ${\sf desc}(\mathcal{C}_0) = \{ {\bf c}_1, {\bf c}_2, {\bf c}_3, {\bf c}_4 \}$,
where ${\bf c}_3 = (a_1, b_1, e_2)^{T}$ and ${\bf c}_4 = (a_1, b_2, e_1)^{T}$.
Then $|{\sf desc}(\mathcal{C}_0) \bigcap \mathcal{C}| \leq 3$.
Otherwise, if $|{\sf desc}(\mathcal{C}_0)  \bigcap  \mathcal{C}| = 4$, i.e.,
${\sf desc}(\mathcal{C}_0) \bigcap \mathcal{C} = \{ {\bf c}_1, {\bf c}_2, {\bf c}_3, {\bf c}_4 \}$,
then ${\sf desc}(\{ {\bf c}_1, {\bf c}_2\}) = {\sf desc}(\{ {\bf c}_3, {\bf c}_4\})$,
a contradiction to the definition of a $\overline{2}$-{\rm SC}.
Since $\mathcal{C}^{'} \subseteq {\sf desc}(\mathcal{C}_0) \bigcap \mathcal{C}$ and $|\mathcal{C}^{'}| \geq 3$,
we have $|\mathcal{C}^{'}| = 3$. So, we may assume, without loss of generality,  that $\mathcal{C}^{'} =  \{ {\bf c}_1, {\bf c}_2, {\bf c}_3\}$.
This implies $\mathcal{C}_0 \subseteq \mathcal{C}^{'}$, a contradiction.

This completes the proof.
\qed

\begin{lemma}
\label{forbiconfi2}
Let $\mathcal{C}$ be a $\overline{2}$-{\rm SC}$(3, M, q)$.
If there exist $\mathcal{C}_0, \mathcal{C}^{'} \subseteq \mathcal{C}$, $|\mathcal{C}_0| \leq 2$, such that
${\sf desc}(\mathcal{C}_0) = {\sf desc}(\mathcal{C}^{'})$ and $\mathcal{C}_0 \not\subseteq  \mathcal{C}^{'}$, then
${\sf desc}(\mathcal{C}_0)\bigcap \mathcal{C}$ is of one of the following four types:
\begin{eqnarray*}
\begin{array}{cccc}
\hbox{Type {\bf I}:} & \hbox{Type {\bf II}:}  \\
\left(
  \begin{array}{cc|cccccc}
    a_1 & a_2 & a_1 & a_1 \\
    b_1 & b_2 & b_1 & b_2 \\
    e_1 & e_2 & e_2 & e_1 \\
  \end{array}
\right),&
\left(
  \begin{array}{cc|cccccc}
    a_1 & a_2 & a_1 & a_2 \\
    b_1 & b_2 & b_1 & b_1 \\
    e_1 & e_2 & e_2 & e_1 \\
  \end{array}
\right),\\[1cm]
\hbox{Type {\bf III:} } & \hbox{Type {\bf IV:}}\\
\left(
  \begin{array}{cc|cccccc}
    a_1 & a_2 & a_1 & a_2 \\
    b_1 & b_2 & b_2 & b_1 \\
    e_1 & e_2 & e_1 & e_1 \\
  \end{array}
\right),&
\left(
  \begin{array}{cc|cccccc}
    a_1 & a_2 & a_1 & a_1 & a_2\\
    b_1 & b_2 & b_1 & b_2 & b_1\\
    e_1 & e_2 & e_2 & e_1 & e_1\\
  \end{array}
\right),
\end{array}
\end{eqnarray*}
where $\mathcal{C}_0 = \{ {\bf c}_1, {\bf c}_2 \}$, ${\bf c}_i = (a_i, b_i, e_i)$, $i = 1, 2$,
and $a_1 \neq a_2$, $b_1 \neq b_2$, $e_1 \neq e_2$.
\end{lemma}
\proof  According to Lemma \ref{forbiconfi1}, we can only have
$\mathcal{C}_0 = \{ {\bf c}_1, {\bf c}_2 \}$, ${\bf c}_i = (a_i, b_i, e_i)^{T}$, where $i = 1, 2$,
$a_1 \neq a_2$, $b_1 \neq b_2$, and $e_1 \neq e_2$.
Then ${\sf desc}(\mathcal{C}_0) = \{ {\bf c}_1, {\bf c}_2, {\bf c}_3, {\bf c}_4, {\bf c}_5, {\bf c}_6, {\bf c}_7, {\bf c}_8 \}$,
where ${\bf c}_3 = (a_1, b_1, e_2)^{T}$, ${\bf c}_4 = (a_1, b_2, e_1)^{T}$, ${\bf c}_5 = (a_2, b_1, e_1)^{T}$,
${\bf c}_6 = (a_2, b_2, e_1)^{T}$, ${\bf c}_7 = (a_2, b_1, e_2)^{T}$, ${\bf c}_8 = (a_1, b_2, e_2)^{T}$. \\
\begin{eqnarray*}
{\bf c}_1 \ \  {\bf c}_2 \ \ \  {\bf c}_3 \ \ \ {\bf c}_4 \ \ \  {\bf c}_5 \ \ \ {\bf c}_6 \ \  \ {\bf c}_7 \ \ \ {\bf c}_8\ \ \ \ \\
 {\sf desc}(\mathcal{C}_0)=
\left(
  \begin{array}{cc|cccccc}
    a_1 & a_2 & a_1 & a_1 & a_2 & a_2 & a_2 & a_1 \\
    b_1 & b_2 & b_1 & b_2 & b_1 & b_2 & b_1 & b_2 \\
    e_1 & e_2 & e_2 & e_1 & e_1 & e_1 & e_2 & e_2 \\
  \end{array}
\right)
\end{eqnarray*}
Let $B_i = \{{\bf c}_{i+2}, {\bf c}_{i+5}\}$, where $1 \leq i \leq 3$.
 Then for any $1 \leq i \leq 3$, we have $B_i \not\subseteq {\sf desc}(\mathcal{C}_0) \bigcap \mathcal{C}$. Otherwise,
${\sf desc}(\mathcal{C}_0) = {\sf desc}(B_i)$,
a contradiction to the definition of a $\overline{2}$-{\rm SC}.
Since $\mathcal{C}$ is a $\overline{2}$-{\rm SC}$(3, M, q)$ and ${\sf desc}(\mathcal{C}_0) = {\sf desc}(\mathcal{C}^{'})$,
we  have $\mathcal{C}^{'} \subseteq {\sf desc}(\mathcal{C}_0)\bigcap \mathcal{C}$ and $|\mathcal{C}^{'}| \geq 3$.

If ${\sf desc}(\mathcal{C}_0) \bigcap \mathcal{C} = \mathcal{C}_0$,
then $\mathcal{C}^{'}  \subseteq \mathcal{C}_0 $,
 and thus $| \mathcal{C}^{'} | \leq | \mathcal{C}_{0} | = 2$, a contradiction.
So ${\sf desc}(\mathcal{C}_0) \bigcap \mathcal{C}$ contains at least one
 of  the words ${\bf c}_3, {\bf c}_4, {\bf c}_5, {\bf c}_6, {\bf c}_7, {\bf c}_8$.
Without loss of generality, we may assume $ {\bf c}_3 \in {\sf desc}(\mathcal{C}_0) \bigcap \mathcal{C}$.
Then ${\bf c}_6 \notin {\sf desc}(\mathcal{C}_0) \bigcap \mathcal{C}$.
If ${\sf desc}(\mathcal{C}_0) \bigcap \mathcal{C} = \{{\bf c}_1, {\bf c}_2, {\bf c}_3\}$,
since $\mathcal{C}^{'} \subseteq {\sf desc}(\mathcal{C}_0) \bigcap \mathcal{C}$ and $|\mathcal{C}^{'}| \geq 3$,
we have $\mathcal{C}^{'} =  \{ {\bf c}_1, {\bf c}_2, {\bf c}_3\}$,
which implies $\mathcal{C}_0 \subseteq \mathcal{C}^{'}$, a contradiction.
So ${\sf desc}(\mathcal{C}_0) \bigcap \mathcal{C}$ should contain at least one
 of the words ${\bf c}_4, {\bf c}_5, {\bf c}_7, {\bf c}_8$.

(1) If ${\bf c}_4 \in {\sf desc}(\mathcal{C}_0) \bigcap \mathcal{C}$, then ${\bf c}_7 \notin {\sf desc}(\mathcal{C}_0) \bigcap \mathcal{C}$.
We also have ${\bf c}_8 \notin {\sf desc}(\mathcal{C}_0) \bigcap \mathcal{C}$, otherwise,
${\sf desc}( \{ {\bf c}_1, {\bf c}_8\})  = {\sf desc}(\{{\bf c}_3, {\bf c}_4\})$, a contradiction.
So, if ${\bf c}_5 \notin {\sf desc}(\mathcal{C}_0) \bigcap \mathcal{C}$,
then ${\sf desc}(\mathcal{C}_0) \bigcap \mathcal{C}$ is of Type {\bf I},
and if ${\bf c}_5 \in {\sf desc}(\mathcal{C}_0) \bigcap \mathcal{C}$,
then ${\sf desc}(\mathcal{C}_0) \bigcap \mathcal{C}$ is of Type {\bf IV}.

(2)  If ${\bf c}_5 \in {\sf desc}(\mathcal{C}_0) \bigcap \mathcal{C}$,
then ${\bf c}_8 \notin {\sf desc}(\mathcal{C}_0) \bigcap \mathcal{C}$.
We also have ${\bf c}_7 \notin {\sf desc}(\mathcal{C}_0) \bigcap \mathcal{C}$, otherwise,
${\sf desc}( \{ {\bf c}_1, {\bf c}_7\}) = {\sf desc}(\{{\bf c}_3, {\bf c}_5\})$, a contradiction.
So, if ${\bf c}_4 \notin {\sf desc}(\mathcal{C}_0) \bigcap \mathcal{C}$,
then ${\sf desc}(\mathcal{C}_0) \bigcap \mathcal{C}$ is of Type {\bf II},
and if ${\bf c}_4 \in {\sf desc}(\mathcal{C}_0) \bigcap \mathcal{C}$,
then ${\sf desc}(\mathcal{C}_0) \bigcap \mathcal{C}$ is of Type {\bf IV}.

(3)  If ${\bf c}_7 \in {\sf desc}(\mathcal{C}_0) \bigcap \mathcal{C}$,
then ${\bf c}_4 \notin {\sf desc}(\mathcal{C}_0) \bigcap \mathcal{C}$.
Also, ${\bf c}_5 \notin {\sf desc}(\mathcal{C}_0) \bigcap \mathcal{C}$, otherwise,
${\sf desc}( \{ {\bf c}_1, {\bf c}_7\}) = {\sf desc}(\{{\bf c}_3, {\bf c}_5\})$, a contradiction.
We further have ${\bf c}_8 \notin {\sf desc}(\mathcal{C}_0) \bigcap \mathcal{C}$,
otherwise, ${\sf desc}( \{ {\bf c}_2, {\bf c}_3\}) = {\sf desc}(\{{\bf c}_7, {\bf c}_8\})$, a contradiction.
So, in this case, ${\sf desc}(\mathcal{C}_0) \bigcap \mathcal{C} = \{ {\bf c}_1, {\bf c}_2, {\bf c}_3, {\bf c}_7\}$.\\
\begin{eqnarray*}
{\bf c}_1 \ \  \ {\bf c}_2 \ \ \ {\bf c}_3 \ \ \ {\bf c}_7\ \ \ \ \ \\
{\sf desc}(\mathcal{C}_0)\bigcap \mathcal{C}=
\left(
  \begin{array}{cc|cccccc}
    a_1 & a_2 & a_1 & a_2 \\
    b_1 & b_2 & b_1 & b_1 \\
    e_1 & e_2 & e_2 & e_2 \\
  \end{array}
\right)
\end{eqnarray*}
If ${\bf c}_1 \notin \mathcal{C}^{'}$ (or ${\bf c}_2 \notin \mathcal{C}^{'}$),
then  $e_1 \notin \mathcal{C}^{'}(3)$ (or $b_2 \notin \mathcal{C}^{'}(2)$),
which implies ${\sf desc}(\mathcal{C}^{'}) \neq {\sf desc}(\mathcal{C}_0)$.
%as $e_1 \in \mathcal{C}_0(3)$(or $b_2 \in \mathcal{C}_0(2)$).
Hence ${\bf c}_1, {\bf c}_2 \in  \mathcal{C}^{'}$,
which implies $\mathcal{C}_0 \subseteq \mathcal{C}^{'}$, a contradiction.
So this case is impossible.

(4)  If ${\bf c}_8 \in {\sf desc}(\mathcal{C}_0) \bigcap \mathcal{C}$,
then ${\bf c}_5 \notin {\sf desc}(\mathcal{C}_0) \bigcap \mathcal{C}$.
Also, ${\bf c}_4 \notin {\sf desc}(\mathcal{C}_0) \bigcap \mathcal{C}$, otherwise,
${\sf desc}( \{ {\bf c}_1, {\bf c}_8\}) = {\sf desc}(\{{\bf c}_3, {\bf c}_4\})$, a contradiction.
We further have ${\bf c}_7 \notin {\sf desc}(\mathcal{C}_0) \bigcap \mathcal{C}$,
otherwise, ${\sf desc}( \{ {\bf c}_2, {\bf c}_3\}) = {\sf desc}(\{{\bf c}_7, {\bf c}_8\})$, a contradiction.
So, in this case, ${\sf desc}(\mathcal{C}_0) \bigcap \mathcal{C} = \{ {\bf c}_1, {\bf c}_2, {\bf c}_3, {\bf c}_8\}$.
\begin{eqnarray*}
{\bf c}_1 \ \  \ {\bf c}_2 \ \  \ {\bf c}_3 \ \ \ {\bf c}_8\ \ \ \ \ \\
{\sf desc}(\mathcal{C}_0)\bigcap \mathcal{C}=
\left(
  \begin{array}{cc|cccccc}
    a_1 & a_2 & a_1 & a_1 \\
    b_1 & b_2 & b_1 & b_2 \\
    e_1 & e_2 & e_2 & e_2 \\
  \end{array}
\right)
\end{eqnarray*}
If ${\bf c}_1 \notin \mathcal{C}^{'}$ (or ${\bf c}_2 \notin \mathcal{C}^{'}$),
then  $e_1 \notin \mathcal{C}^{'}(3)$ (or $a_2 \notin \mathcal{C}^{'}(1)$),
which implies ${\sf desc}(\mathcal{C}^{'}) \neq {\sf desc}(\mathcal{C}_0)$.
%as $e_1 \in \mathcal{C}_0(3)$(or $a_2 \in \mathcal{C}_0(1)$).
Hence ${\bf c}_1, {\bf c}_2 \in  \mathcal{C}^{'}$,
which implies $\mathcal{C}_0 \subseteq \mathcal{C}^{'}$, a contradiction.
So this case is impossible.

This completes the proof.
\qed

\begin{theorem}
\label{resulSSC}
Let $\mathcal{C}$ be a $\overline{2}$-{\rm SC}$(3, M, q)$. Then $\mathcal{C}$ is a $\overline{2}$-{\rm SSC}$(3, M, q)$
if and only if for any $\mathcal{C}_0 = \{ {\bf c}_1, {\bf c}_2 \} =  \{(a_1, b_1, e_1)^{T}$, $(a_2, b_2, e_2)^{T}\}$ $\subseteq \mathcal{C}$, where
$a_1 \neq a_2$, $b_1 \neq b_2$, and $e_1 \neq e_2$,
we have that ${\sf desc}(\mathcal{C}_0)\bigcap \mathcal{C}$ is not of one of the following four types:
\begin{eqnarray*}
\begin{array}{cccc}
\hbox{Type {\bf I}:} & \hbox{Type {\bf II}:}\\
\left(
  \begin{array}{cc|cccccc}
    a_1 & a_2 & a_1 & a_1 \\
    b_1 & b_2 & b_1 & b_2 \\
    e_1 & e_2 & e_2 & e_1 \\
  \end{array}
\right),&
\left(
  \begin{array}{cc|cccccc}
    a_1 & a_2 & a_1 & a_2 \\
    b_1 & b_2 & b_1 & b_1 \\
    e_1 & e_2 & e_2 & e_1 \\
  \end{array}
\right),\\[1cm]
\hbox{Type {\bf III:} } & \hbox{Type {\bf IV:}} \\
\left(
  \begin{array}{cc|cccccc}
    a_1 & a_2 & a_1 & a_2 \\
    b_1 & b_2 & b_2 & b_1 \\
    e_1 & e_2 & e_1 & e_1 \\
  \end{array}
\right),&
\left(
  \begin{array}{cc|cccccc}
    a_1 & a_2 & a_1 & a_1 & a_2\\
    b_1 & b_2 & b_1 & b_2 & b_1\\
    e_1 & e_2 & e_2 & e_1 & e_1\\
  \end{array}
\right),
\end{array}
\end{eqnarray*}
\end{theorem}
\proof Suppose that $\mathcal{C}$ is a $\overline{2}$-{\rm SSC}$(3, M, q)$.
Assume that  there exists $\mathcal{C}_0 = \{ {\bf c}_1, {\bf c}_2 \} =  \{(a_1, b_1, e_1)^{T}, (a_2, b_2,  e_2)^{T}\}  \subseteq \mathcal{C}$,
where $a_1 \neq a_2$, $b_1 \neq b_2$, and $e_1 \neq e_2$, such that
${\sf desc}(\mathcal{C}_0)\bigcap \mathcal{C}$ is of one of the  four types.
For convenience, let ${\bf c}_3 = (a_1, b_1, e_2)^{T}$, ${\bf c}_4  = (a_1, b_2, e_1)^{T}$, ${\bf c}_5  = (a_2, b_1, e_1)^{T}$.

(1)  If ${\sf desc}(\mathcal{C}_0)\bigcap \mathcal{C}$ is  of  type  {\bf I},
then ${\sf desc}(\{ {\bf c}_1, {\bf c}_2\}) = {\sf desc}(\{ {\bf c}_2, {\bf c}_3, {\bf c}_4\} )$,
while $\{ {\bf c}_1, {\bf c}_2\} \not\subseteq \{ {\bf c}_2, {\bf c}_3, {\bf c}_4\}$,
 a contradiction to the definition of a $\overline{2}$-{\rm SSC}. So this case is impossible.

(2)  If ${\sf desc}(\mathcal{C}_0)\bigcap \mathcal{C}$ is  of  type  {\bf II},
then ${\sf desc}(\{ {\bf c}_1, {\bf c}_2\}) = {\sf desc}(\{ {\bf c}_2, {\bf c}_3, {\bf c}_5\} )$,
while $\{ {\bf c}_1, {\bf c}_2\} \not\subseteq \{ {\bf c}_2, {\bf c}_3, {\bf c}_5\}$,
 a contradiction to the definition of a $\overline{2}$-{\rm SSC}. So this case is impossible.

(3)  If ${\sf desc}(\mathcal{C}_0)\bigcap \mathcal{C}$ is  of  type  {\bf III},
then ${\sf desc}(\{ {\bf c}_1, {\bf c}_2\}) = {\sf desc}(\{ {\bf c}_2, {\bf c}_4, {\bf c}_5\} )$,
while $\{ {\bf c}_1, {\bf c}_2\} \not\subseteq \{ {\bf c}_2, {\bf c}_4, {\bf c}_5\}$,
 a contradiction to the definition of a $\overline{2}$-{\rm SSC}. So this case is impossible.

(4)  If ${\sf desc}(\mathcal{C}_0)\bigcap \mathcal{C}$ is  of  type  {\bf IV},
then ${\sf desc}(\{ {\bf c}_1, {\bf c}_2\}) = {\sf desc}(\{ {\bf c}_3, {\bf c}_4, {\bf c}_5\} )$,
while $\{ {\bf c}_1, {\bf c}_2\} \not\subseteq \{ {\bf c}_3, {\bf c}_4, {\bf c}_5\}$,
 a contradiction to the definition of a $\overline{2}$-{\rm SSC}. So this case is impossible.

So, ${\sf desc}(\mathcal{C}_0)\bigcap \mathcal{C}$ is not of one of the  four types described above.

Conversely, suppose that $\mathcal{C}$ is a $\overline{2}$-{\rm SC}$(3, M, q)$,
and for any $\mathcal{C}_0 = \{ {\bf c}_1, {\bf c}_2 \} =  \{(a_1, b_1, e_1)^{T}, \\ (a_2, b_2, e_2)^{T}\} \subseteq \mathcal{C}$,
where $a_1 \neq a_2$, $b_1 \neq b_2$, and $e_1 \neq e_2$,
we have that ${\sf desc}(\mathcal{C}_0)\bigcap \mathcal{C}$ is not of one of the four types.
If $\mathcal{C}$ is not a $\overline{2}$-{\rm SSC}$(3, M, q)$, then there exist
$\mathcal{C}_1 \subseteq \mathcal{C}$, $| \mathcal{C}_1 |\leq 2$, and $\mathcal{C}^{'} \in S(\mathcal{C}_1) =
\{ \mathcal{C}^{'} \subseteq \mathcal{C} \ | \ {\sf desc}(\mathcal{C}^{'}) = {\sf desc}(\mathcal{C}_1)\}$,
such that $\mathcal{C}_1 \not\subseteq \mathcal{C}^{'}$. According
to Lemma \ref{forbiconfi2}, $\mathcal{C}_1 = \{ {\bf c}_1^{'}, {\bf c}_2^{'} \} =
 \{(a_1^{'}, b_1^{'}, e_1^{'})^{T}, (a_2^{'}, b_2^{'},  e_2^{'})^{T}\} \subseteq \mathcal{C}$,
where $a_1^{'} \neq a_2^{'}$, $b_1^{'} \neq b_2^{'}$, and $e_1^{'} \neq e_2^{'}$,
such that ${\sf desc}(\mathcal{C}_1)\bigcap \mathcal{C}$ is of one of the four
types, a contradiction. So $\mathcal{C}$ is a $\overline{2}$-{\rm SSC}$(3, M, q)$.
\qed

Now, we pay our attention to the construction of $\overline{2}$-{\rm SSC}s of length $3$ via the discussion above.

For any $(n,M,q)$ code $\mathcal{C}$ on $Q= \{0,1, \ldots, q-1\}$, Cheng et al. \cite{CJM} defined
the following shortened code ${\mathcal{A}}_{i}^{j}$ for $i \in Q$ and $1 \leq j \leq n$:
\begin{eqnarray*}
 \mathcal{A}_{i}^{j} = \{({\bf c}(1), \ldots, {\bf c}(j-1), {\bf c}(j+1), \ldots, {\bf c}(n))^{T} \ | \
({\bf c}(1), \ldots, {\bf c}(n))^{T} \in \mathcal{C}, {\bf c}(j) = i\}.
\end{eqnarray*}

\begin{lemma}{\rm (\cite{CJM})}
\label{SC}
A $(3, M, q)$ code is a $\overline{2}$-{\rm SC}$(3, M, q)$ on $Q$ if and only if
$|{\mathcal{A}}_{g_1}^{j} \bigcap {\mathcal{A}}_{g_2}^{j}| \leq 1$ holds for any $1 \leq j \leq 3$, and any distinct $g_1, g_2 \in Q$.
\end{lemma}

\begin{lemma}
\label{condiSSC}
Let $\mathcal{C}$ be a  $(3, M, q)$ code on $Q$. If for any $\mathcal{C}_0 \subseteq \mathcal{C}$, $| \mathcal{C}_0 | \leq 2$,
we have $|{\sf desc}(\mathcal{C}_0)\bigcap \mathcal{C}| \leq 3$, then $\mathcal{C}$ is a $\overline{2}$-{\rm SSC}$(3, M, q)$.
\end{lemma}
\proof  We first show that $\mathcal{C}$ is a $\overline{2}$-{\rm SC}$(3, M, q)$.
Assume not. According to Lemma \ref{SC}, we may assume, without loss of generality, that there exist two distinct $g_1, g_2 \in Q$ such that
$|{\mathcal{A}}_{g_1}^{1} \bigcap {\mathcal{A}}_{g_2}^{1}| \geq 2$.
Suppose $(b_1, e_1)^{T}, (b_2, e_2)^{T} \in {\mathcal{A}}_{g_1}^{1} \bigcap {\mathcal{A}}_{g_2}^{1}$, where $(b_1, e_1)^{T} \neq (b_2, e_2)^{T}$.
Then $(g_1, b_1, e_1)^{T}$, $(g_2, b_1, e_1)^{T}$, $(g_1, b_2$, $e_2)^{T}$, $(g_2, b_2, e_2)^{T} \in \mathcal{C}$,
which imply $|{\sf desc}(\{ (g_1, b_1, e_1)^{T}, (g_2, \\b_2, e_2)^{T}\})\bigcap \mathcal{C}| \geq 4$, a contradiction
to the hypothesis. So $\mathcal{C}$ is a $\overline{2}$-{\rm SC}$(3, M, q)$.
Next, we prove it is in fact a $\overline{2}$-{\rm SSC}.
Since for any $\mathcal{C}_0 \subseteq \mathcal{C}$, $| \mathcal{C}_0 | \leq 2$, $|{\sf desc}(\mathcal{C}_0)\bigcap \mathcal{C}| \leq 3$ always holds,
we know that  ${\sf desc}(\mathcal{C}_0)\bigcap \mathcal{C}$ can not be of any of the four types mentioned in Theorem \ref{resulSSC}.
So $\mathcal{C}$ is a $\overline{2}$-{\rm SSC}$(3, M, q)$ from Theorem \ref{resulSSC}.
\qed

In order to describe our construction for $\overline{2}$-{\rm SSC}s of length $3$, we need $s$ new elements
$\infty_i \notin Z_{q-s}$, $i \in \{ 0, 1, \ldots, s-1 \} \subseteq Z_{q-s}$, such that for any $g \in Z_{q-s}$ and any $i \in \{ 0, 1, \ldots, s-1\}$,
\begin{eqnarray*}g + \infty_i = \infty_i + g = g \cdot \infty_i = \infty_i \cdot g = \infty_i.\end{eqnarray*}

Based on Lemma \ref{condiSSC}, we can construct a $\overline{2}$-{\rm SSC}  as follows.

\begin{lemma}
\label{resu2}
Suppose that $q $ is a positive integer, $s$ is a non-negative integer, where $0 \leq s \leq \frac{q}{2}$ and $q-s$ is odd.
Then there exists a  $\overline{2}$-{\rm SSC}$(3, q^2 + sq - 2s^2, q)$.
\end{lemma}
\proof  We will prove this lemma in two steps. That is, we at first construct a $(3, q^2 + sq - 2s^2, q)$ code, and then show that
it is in fact a $\overline{2}$-{\rm SSC}.

Since $q-s$ is odd and $ 0 \leq s \leq \frac{q}{2}$,
we can construct a code $\mathcal{C}$ on $Q = \{ \infty_{0}, \infty_{1}, \ldots,$ $\infty_{s-1}\} \bigcup Z_{q-s}$ as follows. Let
\begin{eqnarray*}  M_{s} =
\left(
  \begin{array}{cccc}
     0   &  0  &  \cdots  &     0   \\
     0   &  1  &  \cdots  &    q-s-1  \\
     0   &  2  &  \cdots  &  2(q-s-1) \\
  \end{array}
\right),
\indent \indent\indent \indent
M_{i} =
\left(
  \begin{array}{ccc}
    \infty_{i} &        i     &     0      \\
        0      &   \infty_{i} &     i      \\
        i      &        0     & \infty_{i} \\
  \end{array}
\right),
\end{eqnarray*}
$i \in \{ 0, 1, \ldots, s-1\}$. Define  ${\mathcal{D}}_j = \{ {\bf c} + g \ | \ {\bf c} \in M_j, g \in Z_{q-s} \}$, where $0 \leq j  \leq s$,
and $\mathcal{C} = \bigcup_{j=0}^{s}{\mathcal{D}}_j$. Then $\mathcal{C}$ is a $(3, q^2 + sq - 2s^2, q)$ code on $Q$.

According to Lemma \ref{condiSSC}, in order to prove that $\mathcal{C}$ is a $\overline{2}$-{\rm SSC}$(3, q^2 + sq - 2s^2, q)$,
it suffices to check that $|{\sf desc}(\mathcal{C}_0)\bigcap \mathcal{C}| \leq 3$ always holds
for any $\mathcal{C}_0 \subseteq \mathcal{C}$, $| \mathcal{C}_0 | \leq 2$. We will check this  in two steps.

(1)  At first, we prove that for any distinct $g_1, g_2 \in Q$,
$(g_1, g_2) \in \{ \infty_{0}, \infty_{1}$, $\ldots, \infty_{s-1}\}^{2} \bigcup  Z_{q-s}^{2}$,
 $|{\mathcal{A}}_{g_1}^{i} \bigcap {\mathcal{A}}_{g_2}^{i}| = 0$
always holds for any  $1 \leq i \leq 3$. We only need  to consider the case
$|{\mathcal{A}}_{g_1}^{1} \bigcap {\mathcal{A}}_{g_2}^{1}| = 0$, because  we can consider the other two cases in a similar way.

(1.1)   For any $0 \leq i< j \leq s-1$, we have $ {\mathcal{A}}_{\infty_{i}}^{1} \bigcap {\mathcal{A}}_{\infty_{j}}^{1} = \emptyset$.
Assume that $(b, e)^{T} \in {\mathcal{A}}_{\infty_{i}}^{1} \bigcap {\mathcal{A}}_{\infty_{j}}^{1}$. Then there exist $b_1, b_2 \in Z_{q-s}$,
such that $(b, e)^{T} = (b_1, b_1 + i)^{T}= (b_2, b_2 + j)^{T}$, which implies $b_1 = b_2 = b$, and $i = j$, a contradiction.

(1.2)   For any distinct $i, j  \in Z_{q-s} $, we have $ {\mathcal{A}}_{i}^{1} \bigcap {\mathcal{A}}_{j}^{1} = \emptyset$.
Assume that $(b, e)^{T} \in {\mathcal{A}}_{i}^{1} \bigcap {\mathcal{A}}_{j}^{1}$.

(1.2.A)   If there exists $ 0 \leq k \leq s-1$ such that $b = \infty_k$,
then $(b, e)^{T} = (\infty_k, i-k)^{T}= (\infty_k, j-k)^{T}$, which implies $i = j$, a contradiction.

(1.2.B)   If there exists $ 0 \leq k \leq s-1$ such that $e = \infty_k$,
then $(b, e)^{T} = (i+k, \infty_k)^{T}= ( j+k,\infty_k)^{T}$, which implies $i = j$, a contradiction.

(1.2.C)   If $b, e \notin \{ \infty_0, \infty_1, \ldots, \infty_{s-1}\}$,
then there exist $b_1, b_2 \in Z_{q-s}$,  such that $(b, e)^{T} = (i+b_1, i+2b_1)^{T}= (j+b_2, j+2b_2)^{T}$.
Hence $i+b_1 = j+b_2$ and $i+2b_1 = j+2b_2$, which imply $b_1 = b_2$ and $i = j$, a contradiction.

(2)   According to (1), we know that  for any distinct $g_1, g_2 \in Q$ and any $1 \leq i \leq 3$,
$| {\mathcal{A}}_{g_1}^{i} \bigcap {\mathcal{A}}_{g_2}^{i} | \geq 1$ implies
$(g_1, g_2) \in  Z_{q-s} \times \{ \infty_0, \infty_1, \ldots, \infty_{s-1}\} $.
We are going to show that $|{\sf desc}(\mathcal{C}_0)\bigcap \mathcal{C}| \leq 3$ always holds
for any $\mathcal{C}_0 \subseteq \mathcal{C}$, $| \mathcal{C}_0 | \leq 2$.
If $| \mathcal{C}_0 | = 1$, then it is clear that
$|{\sf desc}(\mathcal{C}_0)\bigcap \mathcal{C}| = | \mathcal{C}_0 | = 1$.
Now, we consider the case $| \mathcal{C}_0 | = 2$.
Suppose $\mathcal{C}_0 = \{ {\bf c}_1, {\bf c}_2 \} =  \{(a_1, b_1, e_1)^{T}, (a_2, b_2, e_2)^{T}\} \subseteq \mathcal{C}$,
where ${\bf c}_1 \neq {\bf c}_2$. Consider  the Hamming distance of ${\bf c}_1$ and ${\bf c}_2$.

(2.1)   If $d({\bf c}_1, {\bf c}_2) = 1$, then it is clear that
$|{\sf desc}(\mathcal{C}_0)\bigcap \mathcal{C}| = | \mathcal{C}_0 | =2$.

(2.2)   If $d({\bf c}_1, {\bf c}_2) = 2$, without loss of generality, we may assume that $a_1 = a_2$, $b_1 \neq b_2$, $e_1 \neq e_2$.
Then ${\sf desc}(\mathcal{C}_0) = \{ {\bf c}_1, {\bf c}_2, {\bf c}_3, {\bf c}_4 \}$,
where ${\bf c}_3 = (a_1, b_1, e_2)^{T}$ and ${\bf c}_4 = (a_1, b_2, e_1)^{T}$.
\begin{eqnarray*}
{\bf c}_1 \ \  \ {\bf c}_2 \ \  \ {\bf c}_3 \ \ \ {\bf c}_4\ \ \ \ \\
{\sf desc}(\mathcal{C}_0)=
\left(
  \begin{array}{cc|cccccc}
    a_1 & a_1 & a_1 & a_1 \\
    b_1 & b_2 & b_1 & b_2 \\
    e_1 & e_2 & e_2 & e_1 \\
  \end{array}
\right)
\end{eqnarray*}
Assume that $|{\sf desc}(\mathcal{C}_0)\bigcap \mathcal{C}| =  4$, i.e.
${\sf desc}(\mathcal{C}_0)\bigcap \mathcal{C} = \{ {\bf c}_1, {\bf c}_2, {\bf c}_3, {\bf c}_4\}$.
Then $| {\mathcal{A}}_{e_1}^{3} \bigcap {\mathcal{A}}_{e_2}^{3}| \geq 1$,
which implies that exactly one of $e_1$ and $e_2$ is $\infty_i$ for some $0 \leq i \leq s-1$.

(2.2.A)   If $e_1 = \infty_i$, then ${\bf c}_1 ={\bf c}_4$, which implies $|{\sf desc}(\mathcal{C}_0)\bigcap \mathcal{C}| \leq 3$,
a contradiction.

(2.2.B)   If $e_2 = \infty_i$, then ${\bf c}_2 ={\bf c}_3$, which implies $|{\sf desc}(\mathcal{C}_0)\bigcap \mathcal{C}| \leq 3$,
a contradiction.

So, if $d({\bf c}_1, {\bf c}_2) = 2$, $|{\sf desc}(\mathcal{C}_0)\bigcap \mathcal{C}| \leq 3$ always holds.

(2.3)   If $d({\bf c}_1, {\bf c}_2) = 3$, then $a_1 \neq a_2$, $b_1 \neq b_2$, $e_1 \neq e_2$,
and ${\sf desc}(\mathcal{C}_0) = \{ {\bf c}_1, {\bf c}_2, {\bf c}_3, {\bf c}_4, {\bf c}_5, {\bf c}_6, \\ {\bf c}_7, {\bf c}_8 \}$,
where ${\bf c}_3 = (a_1, b_1, e_2)^{T}$, ${\bf c}_4 = (a_1, b_2, e_1)^{T}$, ${\bf c}_5 = (a_2, b_1, e_1)^{T}$,
${\bf c}_6 = (a_2, b_2, e_1)^{T}$, ${\bf c}_7 = (a_2, b_1, e_2)^{T}$, ${\bf c}_8 = (a_1, b_2, e_2)^{T}$.
\begin{eqnarray*}{\bf c}_1 \ \ \  {\bf c}_2 \ \ \ \ {\bf c}_3 \ \ \ {\bf c}_4 \ \ \  {\bf c}_5 \ \ \ {\bf c}_6 \ \   {\bf c}_7 \ \ \ {\bf c}_8\ \ \ \ \ \\
 {\sf desc}(\mathcal{C}_0)=
\left(
  \begin{array}{cc|cccccc}
    a_1 & a_2 & a_1 & a_1 & a_2 & a_2 & a_2 & a_1 \\
    b_1 & b_2 & b_1 & b_2 & b_1 & b_2 & b_1 & b_2 \\
    e_1 & e_2 & e_2 & e_1 & e_1 & e_1 & e_2 & e_2 \\
  \end{array}
\right)
\end{eqnarray*}
We are going to show that ${\sf desc}(\mathcal{C}_0)\bigcap \mathcal{C}$ contains at most one element of the set
$B = \{{\bf c}_3, {\bf c}_4, {\bf c}_5, {\bf c}_6, {\bf c}_7, {\bf c}_8 \}$. Assume not. Then there exist two
elements ${\bf c}^{'}, {\bf c}^{''}$ of $B$ contained in ${\sf desc}(\mathcal{C}_0)\bigcap \mathcal{C}$, where
\begin{eqnarray*}\{ {\bf c}^{'}, {\bf c}^{''} \} \in \{ \{ {\bf c}_3, {\bf c}_4\},
\{{\bf c}_3, {\bf c}_5\}, \{{\bf c}_3, {\bf c}_6\},  \{{\bf c}_3, {\bf c}_7\}, \{{\bf c}_3, {\bf c}_8\},
\{{\bf c}_4, {\bf c}_5\}, \{{\bf c}_4, {\bf c}_6\}, \\ \{{\bf c}_4, {\bf c}_7\},
\{{\bf c}_4, {\bf c}_8\}, \{{\bf c}_5, {\bf c}_6\}, \{{\bf c}_5, {\bf c}_7\},
\{{\bf c}_5, {\bf c}_8\}, \{{\bf c}_6, {\bf c}_7\},
\{{\bf c}_6, {\bf c}_8\}, \{{\bf c}_7, {\bf c}_8\}\}.\end{eqnarray*} However, we can
prove none of them is possible.

(2.3.A)   If $\{{\bf c}^{'}, {\bf c}^{''} \} = \{ {\bf c}_3, {\bf c}_4\}$, then we have
$\{ {\bf c}_1, {\bf c}_2, {\bf c}_3, {\bf c}_4\} \subseteq  {\sf desc}(\mathcal{C}_0)\bigcap \mathcal{C} $.
\begin{eqnarray*}{\bf c}_1 \ \  \ {\bf c}_2 \ \ \ {\bf c}_3 \ \ \ {\bf c}_4 \ \ \ \ \\
\left(
  \begin{array}{cc|cccccc}
    a_1 & a_2 & a_1 & a_1 \\
    b_1 & b_2 & b_1 & b_2 \\
    e_1 & e_2 & e_2 & e_1 \\
  \end{array}
\right)
\end{eqnarray*}
Then $| {\mathcal{A}}_{e_1}^{3} \bigcap {\mathcal{A}}_{e_2}^{3}| \geq 1$ (from ${\bf c}_1$ and ${\bf c}_3$) and
$| {\mathcal{A}}_{b_1}^{2} \bigcap {\mathcal{A}}_{b_2}^{2}| \geq 1$ (from ${\bf c}_1$ and ${\bf c}_4$). Hence
there exist $0 \leq i , j\leq s-1$ such that exactly one of $e_1$ and $e_2$ is $\infty_i$, and
exactly one of $b_1$ and $b_2$ is $\infty_j$.

(2.3.A.a)   If $e_1 = \infty_i$, then $\infty_j \notin \{b_1, b_2\}$ from ${\bf c}_1$ and ${\bf c}_4$, a contradiction.
So, this case is impossible.

(2.3.A.b)   If $e_2 = \infty_i$, then $\infty_j \notin \{b_1, b_2\}$ from ${\bf c}_2$ and ${\bf c}_3$, a contradiction.
So, this case is impossible.

Similarly, we can know that it is impossible for $$\{ {\bf c}^{'}, {\bf c}^{''} \} \in \{
\{{\bf c}_3, {\bf c}_5\},  \{{\bf c}_4, {\bf c}_5\}, \{{\bf c}_6, {\bf c}_7\},
\{{\bf c}_6, \\ {\bf c}_8\}, \{{\bf c}_7, {\bf c}_8\}\}.$$

(2.3.B)   If $\{{\bf c}^{'}, {\bf c}^{''} \} = \{ {\bf c}_3, {\bf c}_6\}$, then we have
$\{ {\bf c}_1, {\bf c}_2, {\bf c}_3, {\bf c}_6\} \subseteq  {\sf desc}(\mathcal{C}_0)\bigcap \mathcal{C} $.
\begin{eqnarray*}{\bf c}_1 \ \  \ {\bf c}_2 \ \ \  {\bf c}_3 \ \ \ {\bf c}_6 \ \ \ \ \\
\left(
  \begin{array}{cc|cccccc}
    a_1 & a_2 & a_1 & a_2 \\
    b_1 & b_2 & b_1 & b_2 \\
    e_1 & e_2 & e_2 & e_1 \\
  \end{array}
\right)
\end{eqnarray*}
Then $| {\mathcal{A}}_{e_1}^{3} \bigcap {\mathcal{A}}_{e_2}^{3}| \geq 1$ (from ${\bf c}_1$ and ${\bf c}_3$). Hence, without loss of generality,
we may assume that $e_1 \in Z_{q-s}$ and  there exists $0 \leq i \leq s-1$ such that $e_2 =\infty_i$.
Then we can derive that $a_1, a_2, b_1 = a_1 + i, b_2 = a_2 + i \in Z_{q-s}$, which imply ${\bf c}_1, {\bf c}_6 \in {\mathcal{D}}_s$.
So we can derive $e_1 = a_1 + 2i = a_2 + 2i$, which implies $a_1 = a_2$, a contradiction.
So this case is impossible.

Similarly, it is impossible that $\{ {\bf c}^{'}, {\bf c}^{''} \} \in \{
\{{\bf c}_4, {\bf c}_7\},  \{{\bf c}_5, {\bf c}_8\} \}$.

(2.3.C)   If $\{{\bf c}^{'}, {\bf c}^{''} \} = \{ {\bf c}_3, {\bf c}_7\}$, then we have
$\{ {\bf c}_1, {\bf c}_2, {\bf c}_3, {\bf c}_7\} \subseteq  {\sf desc}(\mathcal{C}_0)\bigcap \mathcal{C} $.
\begin{eqnarray*}{\bf c}_1 \ \  \ {\bf c}_2 \ \ \  {\bf c}_3 \ \ \ {\bf c}_7 \ \ \ \   \\
\left(
  \begin{array}{cc|cccccc}
    a_1 & a_2 & a_1 & a_2 \\
    b_1 & b_2 & b_1 & b_1 \\
    e_1 & e_2 & e_2 & e_2 \\
  \end{array}
\right)
\end{eqnarray*}
Then $| {\mathcal{A}}_{e_1}^{3} \bigcap {\mathcal{A}}_{e_2}^{3}| \geq 1$ (from ${\bf c}_1$ and ${\bf c}_3$),
$| {\mathcal{A}}_{b_1}^{2} \bigcap {\mathcal{A}}_{b_2}^{2}| \geq 1$ (from ${\bf c}_2$ and ${\bf c}_7$),
and $| {\mathcal{A}}_{a_1}^{1} \bigcap {\mathcal{A}}_{a_2}^{1}| \geq 1$ (from ${\bf c}_3$ and ${\bf c}_7$).
Hence there exists $0 \leq i , j, k \leq s-1$ such that exactly one of $e_1$ and $e_2$ is $\infty_i$, and $\infty_j \in \{b_1, b_2\}$,
$\infty_k \in \{a_1, a_2\}$. Then at least one of $(a_1, b_1, e_1)^{T}$ and $(a_2, b_2, e_2)^{T}$ contains
at least two components from $\{ \infty_{0}, \infty_{1}, \ldots, \infty_{s-1}\}$, a contradiction. So this case is impossible.

Similarly, it is impossible that
$$\{ {\bf c}^{'}, {\bf c}^{''} \} \in \{ \{{\bf c}_3, {\bf c}_8\},
 \{{\bf c}_4, {\bf c}_6\},  \{{\bf c}_4, {\bf c}_8\}, \{{\bf c}_5, {\bf c}_6\},
\{{\bf c}_5, {\bf c}_7\}\}.$$

The conclusion then comes from Lemma \ref{condiSSC}.
\qed

\begin{theorem}
\label{mainresu}
There exists a  $\overline{2}$-{\rm SSC}$(3, \frac{1}{8}(9q^2 - w^2), q)$ for any positive integer $q$,
with $m$ being the residue of  $q$ modulo $8$, and
\begin{eqnarray*}w =
\left\{\begin{array}{rl}
4-m,  \ \ \ \ \ \ \  \ \ \ \  &   \mbox{if}  \ \ m \equiv 0 \pmod 4,\\[2pt]
\min\{m, 8 - m\},  &  \mbox{otherwise}.\\[2pt]
\end{array}
\right.
\end{eqnarray*}
\end{theorem}
\proof  According to Lemma \ref{resu2}, there exists a  $\overline{2}$-{\rm SSC}$(3, q^2 + sq - 2s^2, q)$ for any positive integer $q$,
where $ 0 \leq s \leq \frac{q}{2}$,
and $q - s$ is odd. Let $q = 8r + m$, where $r$ is a non-negative integer, and  $f(s) = q^2 + sq - 2s^2 = -2(s-\frac{q}{4})^2 + \frac{9}{8}q^2$.
Now, we  are going to find the maximum value of $f(s)$, where $0 \leq s \leq \frac{q}{2}$
and $q - s$ is odd.

(1)   If $m =0$, then $q$ is even. Since $\frac{q}{4} = 2r$ is even, $s = 2r - 1 = \frac{q - 4}{4}$ is odd, then  $q-s$ is odd, and
$f(\frac{q - 4}{4}) = \frac{1}{8}(9q^2 -4^2)$ is the maximum value of $f(s)$.

(2)   If $m =1$, then $q$ is odd and $\frac{q}{4} = 2r + \frac{1}{4}$.
Since $s = 2r  = \frac{q -1}{4}$ is even,  then $q-s$ is odd, and
$f(\frac{q -1}{4}) = \frac{1}{8}(9q^2 - 1)$ is the maximum value of $f(s)$.

(3)   If $m =2$, then $q$ is even and $\frac{q}{4} = 2r + \frac{2}{4}$.
Since $s = 2r + 1 = \frac{q+2}{4}$ is odd, then $q-s$ is odd, and
$f(\frac{q+2}{4})  = \frac{1}{8}(9q^2 -2^2)$ is the maximum value of $f(s)$.

(4)   If $m = 3$, then $q$ is odd and $\frac{q}{4} = 2r + \frac{3}{4}$.
Since $s = 2r  = \frac{q - 3}{4}$ is even, then $q-s$ is odd, and
$f(\frac{q - 3}{4})  = \frac{1}{8}(9q^2 - 3^2)$ is the maximum value of $f(s)$.

(5)   If $m = 4$, then $q$ is even.  Since
$s = 2r + 1 = \frac{q}{4}$ is odd, then  $q-s$ is odd, and
$f(\frac{q}{4})   = \frac{9}{8}q^2$ is the maximum value of $f(s)$.

(6)   If $m = 5$, then $q$ is odd and $\frac{q}{4} = 2r + \frac{5}{4}$.
Since $s = 2r + 2 = \frac{q + 3}{4}$ is even, then  $q-s$ is odd, and
$f(\frac{q + 3}{4})  = \frac{1}{8}(9q^2 - 3^2)$ is the maximum value of $f(s)$.

(7)   If $m = 6$, then $q$ is even and $\frac{q}{4} = 2r + \frac{6}{4}$.
Since $s = 2r + 1 = \frac{q - 2}{4}$ is odd, then $q-s$ is odd, and
$f(\frac{q - 2}{4})  = \frac{1}{8}(9q^2 - 2^2)$ is the maximum value of $f(s)$.

(8)   If $m = 7$, then $q$ is odd and $\frac{q}{4} = 2r + \frac{7}{4}$.
Since $s = 2r + 2 = \frac{q + 1}{4}$ is even, then $q-s$ is odd, and
$f(\frac{q + 1}{4})  = \frac{1}{8}(9q^2 - 1)$ is the maximum value of $f(s)$.

We can summarize the results obtained in (1)-(8) into the following table, from which the conclusion comes.

\begin{center}
\begin{tabular}{|c|c|c|c|}
  \hline
  $ m $ & $ w $ &          $ s $            & $f(s)$ \\ \hline
  $ 0 $ & $ 4 $ & $ \frac{1}{4}(q - 4) $     & $ \frac{1}{8}(9q^2 -4^2) $ \\ \hline
  $ 1 $ & $ 1 $ & $ \frac{1}{4}(q - 1) $     & $ \frac{1}{8}(9q^2 -1^2) $ \\ \hline
  $ 2 $ & $ 2 $ & $ \frac{1}{4}(q + 2) $     & $ \frac{1}{8}(9q^2 -2^2) $ \\ \hline
  $ 3 $ & $ 3 $ & $ \frac{1}{4}(q - 3) $     & $ \frac{1}{8}(9q^2 -3^2) $ \\ \hline
  $ 4 $ & $ 0 $ & $ \frac{q}{4}     $     & $ \frac{9}{8}q^2         $ \\ \hline
  $ 5 $ & $ 3 $ & $ \frac{1}{4}(q + 3) $     & $ \frac{1}{8}(9q^2 -3^2) $ \\ \hline
  $ 6 $ & $ 2 $ & $ \frac{1}{4}(q - 2)$     & $ \frac{1}{8}(9q^2 -2^2) $ \\ \hline
  $ 7 $ & $ 1 $ & $ \frac{1}{4}(q + 1) $     & $ \frac{1}{8}(9q^2 -1^2) $ \\ \hline
\end{tabular}
\end{center}

%\begin{center}
%\begin{tabular}{|c|c|c|c|c|c|c|c|c|c|c|c|c|c|c|c|c|c|}
  %\hline
  %$ q $                                        &  $4$ & $12$ & $20$  & $28$  & $36$  & $44$  % & $52$  & $60$
                                               %& $68$  & $76$  & $84$ & $92$ & $100$  \\ \hline
  %$2$-FPC$(3, q^2, q)$                         &  $16$ & $144$ & $400$  & $784$  & $1296$
                                               %& $1936$  & $2704$  & $3600$ & $4624$ & $5776$  & $7056$  & $8464$  & $10000$\\ \hline
  %$\overline{2}$-SSC$(3, \frac{9}{8}q^2 , q)$  & $ 1 $ & $ \frac{1}{4}(q - 1) $    \\ \hline

%\end{tabular}
%\end{center}

\qed

As is well-known, for any $2$-FPC$(3, M, q)$, we have $M \leq q^2$ (see for example \cite{BT}).
Theorem \ref{mainresu} shows that there is an infinite series of $\overline{2}$-{\rm SSC}$(3, M^{'}, q)$s
which have more than $12.5\%$ codewords than $2$-FPC$(3, M, q)$s could have.
For example, we compare the number of codewords between $2$-FPC$(3, q^2, q)$ and $\overline{2}$-SSC$(3, \frac{9}{8}q^2, q)$,
where $q \equiv 4 \pmod 8$ and $ 4 \leq q \leq 100$.

\vskip24pt

\begin{center}
\begin{tabular}{|c|c|c|}
  \hline
  $ q $ &  $q^2$ (FPC) & $\frac{9}{8}q^2$ (SSC)  \\ \hline    %  & $f(s)$ \\ \hline
  $4   $   & $ 16   $    & $ 18   $   \\ \hline
  $12  $   & $ 144  $    & $ 162  $    \\ \hline
  $ 20 $   & $ 400  $    & $ 450  $     \\ \hline
  $ 28 $   & $ 784  $    & $ 882  $    \\ \hline
  $ 36 $   & $ 1296 $    & $ 1458 $      \\ \hline
  $ 44 $   & $ 1936 $    & $ 2178 $    \\ \hline
  $ 52 $   & $ 2704 $    & $ 3042 $    \\ \hline
  $ 60 $   & $ 3600 $    & $ 4050 $    \\ \hline
  $ 68 $   & $ 4624 $    & $ 5202 $    \\ \hline
  $ 76 $   & $ 5776 $    & $ 6498 $    \\ \hline
  $ 84 $   & $ 7056 $    & $ 7938 $    \\ \hline
  $ 92 $   & $ 8464 $    & $ 9522 $    \\ \hline
  $ 100$   & $ 10000$    & $ 11250$    \\ \hline
\end{tabular}
\end{center}

\vskip24pt

\section{Conclusions}
\label{conclu}

In this paper, we first introduced a new notion of an SSC  for multimedia fingerprinting to resist the averaging attack,
and considered its tracing algorithm to identify colluders.  We showed that SSCs are weaker than frameproof codes,
and can be used to  identify all colluders with computational complexity linear to the product of the length of the code and
the number of authorized users.
We also obtained several infinite series of optimal $\overline{2}$-SSCs of length $2$, and constructed
an infinite series of $\overline{2}$-SSCs of length $3$ with many codewords.

\end{document}